\documentstyle{laa}      % LaTeX A&A Standard Fonts

\begin{document}
\thesaurus{03        % A&A Section 3: Extragal. Astr.
           (19.92.1  % Supernovae and supernova remnants: general
            19.94.1  % Surveys
            07.09.1  % Galaxies: general
            07.22.1) % Galaxies: stellar content of
                   }
\title{The rate of supernovae}

\subtitle{II. The selection effects and the frequencies per unit blue
luminosity}

\author{E.\ Cappellaro\inst{1} \and M. Turatto\inst{1}
\and S.\ Benetti\inst{2} \and D.Yu. Tsvetkov\inst{3}
\and  O.S. Bartunov\inst{3} \and I.N. Makarova\inst{3}}

\offprints{E. Cappellaro}

\institute {Osservatorio Astronomico di Padova, vicolo dell'Osservatorio 5,
I-35122 Padova, Italy \and
Dipartimento di Astronomia, Universit\'a of Padova, vicolo dell'Osservatorio 5,
I-35122 Padova, Italy \and
Sternberg Astronomical Institute, Universitetskij Prospect V-234,
119899 Moscow, Russia}

\date{Received ................; accepted ................}

\maketitle

\begin{abstract}
We present new estimates of the observed rates of SNe determined with the {\em
control time} method applied to the files of observations of two long term,
photographic SN searches carried out at the Asiago and Sternberg Observatories.
Our calculations are applied to a galaxy sample extracted from RC3, in which 65
SNe have been discovered. This relatively large number of SNe has been
redistributed in the different morphological classes of host galaxies giving
the respective SN rates.

The magnitude of two biases, the overexposure of the central part of galaxies
and the inclination of the spiral parent galaxies, have been estimated. We show
that due to overexposure a increasing fraction of SNe is lost in galaxies of
increasing distances. Also, a reduced number of SNe is discovered in inclined
galaxies ($i>30\degr$): SNII and Ib are more affected than Ia, as well as SNe
in Sbc-Sd galaxies with respect to other spirals.

We strengthen previous findings that the SN rates is proportional to the galaxy
blue luminosity for all SN and Hubble types.

Other sources of errors, besides those due to the statistics of the events,
have been investigated. In particular those related to the adopted SN
parameters (Cappellaro et al. (\cite{paper1})) and correction factor for
overexposure and inclination. Moreover, we show that the frequencies of SNe
per unit luminosity vary if different sources for the parameters of the sample
galaxies are adopted, thus hampering the comparison of SN rates based on
different galaxy samples.

The overall rates per unit blue luminosity are similar to the previous
determinations but significant differences show up for individual types. In
particular, the rate in ellipticals, $0.11$ SNu ($H=75\,km\,s^{-1}\,Mpc^{-1}$),
is significantly lower than previously reported and better agrees with the
predictions of galaxy evolutionary models. Contrary to recent claims, in late
spirals the rates of SNIa ($0.39$ SNu) and Ib ($0.27$ SNu) are similar.
The most frequent SNe in spiral galaxies are SNII ($1.48$ SNu).
Even the possible occurrence of faint SNe similar to SN 1987A ($<0.5$
SNu in late spirals) does not significantly alter the total rate of SNII.

In the Galaxy, the expected number of SNe is $1.7\pm0.9$ per century.

\end{abstract}

\section{Introduction}
In the past, two different approaches have been followed to estimate the rates
of supernovae (SNe) in external galaxies. Such calculation were either based on
all the SNe discovered in a given galaxy sample or they were restricted to the
SNe discovered in a single search program. In the former case it was necessary
to make some assumption on the overall surveillance of the galaxy sample (e.g.
Tammann \cite{tammann:82}) whereas, in the latter case, it was possible to
compute accurately the control time for each individual galaxy from the log of
observations of the search (e.g. Cappellaro \& Turatto \cite{capp:tur}
hereafter CT88, Evans et al. \cite{evans:etal}.

Regardless of the method used, the comparison between recently published
estimates of the SN rates shows  general agreement on the gross numbers, but
highlights the uncertainty of the rate for a single SN type and/or in a given
galaxy type (for a recent review see van den Bergh and Tammann
\cite{vdb:tamm}, hereafter vdB\&T).

The main purpose of the present work is to reduce the problem of small number
statistics by combining two independent SN searches conducted, for
almost three decades, at the Asiago Astrophysical Observatory (Italy) and at
the Crimean Station of the Sternberg Institute of Moscow (Russia).
Also, we improve previous calculations allowing for a more detailed SN
taxonomy: SN rates are estimated independently for SN Ia, Ib/c and, when the
statistic allows it, for IIP (Plateau) and IIL (Linear). An upper limit is
given also for faint SN~II similar to 1987A.

In Paper I (Cappellaro et al. \cite{paper1}) we presented the SN and the galaxy
samples on which we base the new estimates. In that paper we also analysed the
computational recipe and the errors induced by the uncertainties in the
parameters involved. In particular, we stressed the relatively strong influence
of the uncertainties in the limiting discovery magnitudes, on the average
absolute SN magnitudes and on the light curves shape. The effect due
to the dispersion of the absolute SN magnitudes turned out to be negligible.
Finally, we emphasized that in comparing SN rate estimates of different
authors one must account for the adoption of a particular galaxy catalogue. In
fact, galaxy catalogues differ not only because of their selection criteria,
but also in the description of individual galaxies. Particularly important for
the SN rates are differences/errors in the galaxy morphological classification
and, as we will show in this paper, in the galaxy luminosity.

\section{RC3 galaxy sample}
After completion of Paper I, the Third Reference Catalogue of Bright Galaxies
of de Vaucouleurs et al. (\cite{rc3}, hereafter RC3) became available. The RC3
is expected to be reasonably complete for galaxies with angular diameter
$D_{25}>1\farcm 0$, total magnitude $B_{\rm T}<15.5$ and recession velocity
$v<15000\,km\,s^{-1}$ ($\sim 12000$ galaxies). This, combined with the relevant
information reported for each galaxy, makes this catalogue particularly
suitable for our investigation.

Following the recipe illustrated in Paper I, we first selected the RC3 galaxies
which appear in the fields of our surveys (4301 galaxies). We then retained
only those galaxies for which the redshift, the morphological type, the
axial ratio and the $B_{\rm T}$ magnitude are available (2461 galaxies). The
total control times ($ICT = \sum_{i=1}^{n} tct_i$, where $tct_i$ is the total
control time of the $i$-th galaxy and $n$ is the number of galaxies) for the
different SN types  of the RC3 galaxy sample are: ICT(Ia)=17270, ICT(Ib)=11510,
ICT(IIP)=10820 and ICT(IIL) = 9783 years (cf. Table 6 of Paper I). In the RC3
galaxy sample, using to the prescription discussed Paper I, we selected
65 SNe (cf. Tab~2 of Paper I).

Note that while the number of galaxies in the RC3 sample is about 50\% larger
than it was for the RC2 sample it yielded the same number of SNe. The main
reason is that the RC3 sample contains a significant fraction of distant
galaxies with small control times, each of which gives a small contribution to
the SN statistics. In fact the ICTs of the RC3 are only $\sim15\%$ larger than
for RC2 sample.

The reader is reminded that RC2 included all SN parent galaxies, known at the
time of publication. Therefore it cannot be considered an unbiased collection
of
galaxies for SN statistics. In principle this is also reflected in RC3, since
it includes all RC2 galaxies. However, a quick investigation shows that only
one of the parent galaxies of our selection (NGC 4525) fails the limits of
completeness of the RC3, because of its  small diameter. Therefore, we can
safely assume that the RC3 galaxy sample is not biased in favour of SNe
producers.

\section{Selection effects on the SN discovery}\label{corr:sec}
The control time of a galaxy is the time during which a given SN stays above
the detection threshold. Sometimes a SN may be lost, even if brighter than the
assumed limiting discovery magnitude $m_{\rm lim}$, because of various
random or systematic effects. Random effects include plate defects, poor
weather conditions, errors of the hunter, etc. We showed in Paper I that,
because of the oversampling of our combined survey, this kind of errors has
only a minor impact on the calculated SN rates ($<3\%$).

Systematic effects such as the discovery of SNe in the brightest regions of
galaxies and in inclined galaxies result more severe bias. Overexposure, in
particular, is expected to change with the scale of the telescope, the exposure
time, etc.. Hence it is expected that the {\em correction factor} will depend
on
the different types of instrumentation. In the following, because of the
similarity of the Asiago and Crimea searches (both based on wide field
telescopes and photographic plates and reaching similar limiting magnitude), we
will analyse these effects on the combined surveys.

\subsection{Overexposure of the nuclear regions of galaxies} \label{secover}
It has been shown by Shaw (\cite{shaw}) that, in SN searches, a number of SNe
in the central regions of the galaxies is lost. This effect was found to be
dependent on the distance of the host galaxy and is relatively strong in
photographic surveys. This is so because the central regions of such galaxies
are overexposed due to the low dynamic range of the plates (Bartunov et al.
\cite{bart:etal}). Visual and CCD surveys appear less affected by this effect.

To evaluate the size of this effect, we compared the radial distribution of SNe
discovered in our combined search with that of the SNe discovered visually by
Evans (Evans et al. \cite{evans:etal}) and by the CCD Berkeley automated
supernova search (hereafter BASS, Muller et al. \cite{muller:etal}) considered
together. For each SN the projected relative distance from the nucleus was
calculated using the relation: \\
\centerline{$r = 2 \times (d_a^2 + d_b^2)^{1/2}/D_{25}$ } \\
where $d_a$ and $d_b$ are the measured offset of the SN from the nucleus and
$D_{25}$ is the apparent diameter of the parent galaxy.  Since here we are
interested only to the locations of SNe within the host galaxies (not in the
rates), we can use all SNe in Tab.2 of Paper I whose parent galaxy distances
are known. The radial distribution of SNe in galaxies at different recession
velocities are listed in Table~\ref{radrel}.

\begin{table*}
\caption{Radial distribution of SNe in galaxies at different distances}
\label{radrel}
\begin{tabular}{ccccccc}
\hline
&\multicolumn{3}{c}{Asiago+Crimea}&&\multicolumn{2}{c}{Evans+BASS}\\
\cline{2-4}\cline{6-7}
$r/R$  & $v\le2000$ & $2000<v\le4000$ & $v>4000$&& $v\le2000$ &
$2000<v\le4000$\\
\hline
$\le 0.25$     & 12  &  4   &   2   &&  12   &   4\\
$0.25 to 0.75$& 24  & 11   &  15   &&  14   &  10\\
$>0.75$        &  6  &  2   &   3   &&   3   &   2\\
               &     &      &       &&       &    \\
Lost SNe       & 18\%  & 23\%   &  35\% & &  **   &  22\%\\
$f_{\rm v}$    &0.82  & 0.77 & 0.65     & &         &      \\
\hline
\end{tabular}

** Adopted reference distribution.
\end{table*}

First of all let us compare the radial distributions in the two groups of
searches (the Asiago+Crimea versus the Evans+BASS) for SNe that exploded in
nearby galaxies ($v < 2000\,km\,s^{-1}$). Normalizing the two distributions to
the outer regions ($r>0.25$) in the photographic surveys results in a
deficiency
of SNe in the inner regions. Assuming that the visual/CCD searches did not miss
SNe in the central regions of nearby galaxies we find that our searches lost
$\simeq18\%$ of all SNe.

We then compared the radial distributions of SNe in galaxies of increasing
redshift in order to test the finding by Shaw (\cite{shaw}) that increasing the
distance the effect is larger. It turns out that for parent galaxies in the
range 2000 to 4000 $km\,s^{-1}$ our photographic searches  lost $\simeq 23\%$
of the SNe, whereas for the more distant galaxies ($v>4000\,km\,s^{-1}$) we
lost about 35\% of SNe (relative to the nearby Evans+BASS sample). Finally we
note that, contrary to previous claims,  even the combined Evans+BASS sample is
affected by this bias for the more distant galaxies (Tab.~\ref{radrel}). They
lost $\simeq 22\%$ of all SNe in parent galaxies in the range 2000 to 4000
$km\,s^{-1}$ (where the BASS contribution is dominant).

At first sight, the effect appears significantly smaller than that derived by
Shaw (\cite{shaw}, Fig. 3). However, this in mainly due to the different
units adopted for the radial distances of SNe. If we carry out the same
exercise using for each SN the absolute radial distance (in kpc) instead of the
relative distance as in Tab.~\ref{radrel}, we find that 36\% of the SNe
exploded in galaxies between 2000 and 4000$\,km\,s^{-1}$ were lost, whereas for
$v>4000\,km\,s^{-1}$ this percentage is 49\%. These numbers are in excellent
agreement with Shaw (\cite{shaw}).
However, one must keep in mind that, at large distance, the galaxy sample is
biased in favour of intrinsically bright and large galaxies. Therefore SNe in
distant galaxies are expected to have (in absolute units) a less peaked radial
distribution. This is not an observational selection effect, but rather it
results from the intrinsic properties of the galaxy selection. We prefer,
therefore, to use the correction factors derived from the relative radial
distributions.

Unfortunately, given the small numbers, it was not possible to check the
possible variations of this bias for the different SN types and galaxy
morphological types. In principle such a dependence is expected because of the
different surface brightness profile of the various galaxy types and because of
the different stellar populations to which various SN types belong.

The net result of overexposure is an underestimate of the SN rates. We
account for it by simply multiplying the control time of each individual galaxy
by an average factor, $f_{\rm v}$, which is function of the galaxy distance as
shown in the last row of Table~\ref{radrel}.

\subsection{Inclination of spiral galaxies} \label{secinc}
The probability of discovering SNe in spiral galaxies depends on the
inclination of the disk relative to the line of sight; SNe in more
inclined galaxies being more difficult to find. At least for photographic
surveys, this effect has been verified by several authors, even if different
correction factors were found in different samples (Tammann \cite{tammann:77},
CT88, Guthrie \cite{gut}, van den Bergh \& McClure \cite{vdb:mcc}).
Usually it is assumed that the cause of this bias is the presence of
absorbing layers close to the disks of spiral galaxies, that probably have been
hollowed out by multiple SN explosions (van den Bergh \& McClure
\cite{vdb:mcc}). Alternatively, it has been suggested that the higher surface
brightness of inclined spirals cause the overexposure of
disks on photographic plates. This would explain why this selection
effect is found not to be so important for visual (vdB\&T) or CCD (Muller et
al.
\cite{muller:etal}) surveys.

\begin{figure*}
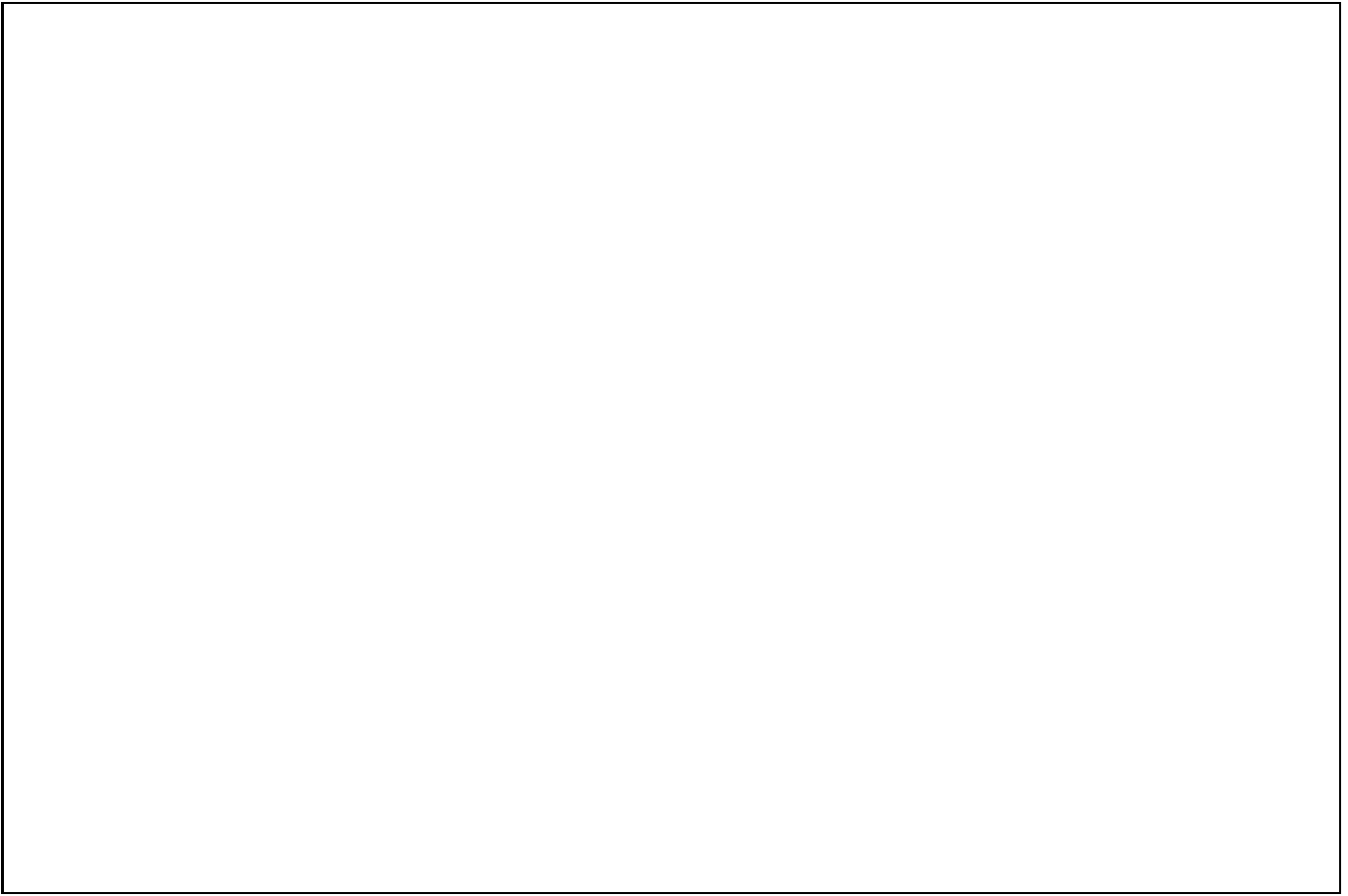

\picplace{12cm}
\caption[]{Effect of galaxy inclination on the SN rate. In the upper left panel
the whole sample of SNe in spirals is considered, while on the two right panels
the overall SN rate is reported, separately, for early and late spirals. Due to
the small statistics, here, SN parent galaxies have been split only in two
inclination bins, $0\degr$ to $30\degr$ and $30\degr$ to $90\degr$. Lower
panels
show the SN rates in the whole sample of spirals for SNe Ia, Ib, II. In
brackets are the numbers of SNe in each sample. Indicative error bars were
computed assuming a Poisson statistics for the SN events. }
\label{inc} \end{figure*}

Regardless of the causes, we try to estimate the size of this bias in our
searches for the galaxies of the RC3 sample. First, we computed the
inclination angle $i$ along the line of sight, from the isophotal axial ratio
$d/D$, and through the  relation (it is assumed that an edge--on system would
be measured to have the axial ratio $d/D=0.2$):
$$i = \cos^{-1}\{[(d/D)^2-0.2^2]/[1-0.2^2]\}^{1/2}$$
Then, the whole spiral galaxy sample was divided in three inclination bins and
the SN rates per average galaxy were calculated in each bin. Note that for
this purpose the control times have been corrected for the overexposure effect
as determined in Sect. \ref{secover}. The results for the different types of
SNe and galaxies are shown in Fig.~\ref{inc}.

Even if the statistics are relatively poor, expecially when considering only
a particular type of SNe, a number of useful indications can be obtained from
Fig. \ref{inc}:

\begin{enumerate}
\item The inclination bias is important: in the whole sample the rate of SNe
shows a sharp peak in face--on galaxies; the {\em observed} SN rate in inclined
 galaxies ($i>30\degr$) being about 3 times smaller. A similar result was found
by van den Bergh and McClure (\cite{vdb:mcc}), who argued in favour of chimney
like structures in the absorbing layers.

\item The bias is more severe in the late than in the early spirals, as found
also by van den Bergh and McClure (\cite{vdb:mcc}) and by Guthrie (\cite{gut}).
Whereas in the inclined early spirals (S0a-Sb) the factor is only $\sim 2$, in
the late spirals (Sbc-Sd) the rate of SNe is depressed by a factor $\sim 4$
(possibly only due to SNeII).

\item The bias appears dependent on the type of SNe considered: it is
stronger for Ib and II (factor $\sim$4), smaller for Ia ($\sim$1.5). Among
SNeII, it appears stronger for IIP than for IIL but, due to the poor
statistics, we will neglect this tentative result.
\end{enumerate}

For Sdm-Sm galaxies, the statistics are very poor since only two SNe in our
sample belong to parent galaxies of this type. To gain some indication,
in analogy to van den Bergh \& McClure (\cite{vdb:mcc}), we compared the
distribution of the inclinations of all Sdm-Sm parent galaxies listed in the
Asiago SN Catalogue (Barbon et al. \cite{cat}) with that of the RC3
galaxies of this type. It appears that the effect in Sdm-Sm is not so strong as
for Sbc-Sd and, therefore, we applied to Sdm-Sm the same factors as for S0a-Sb.

In order to correct for this selection effect, the control time
of each inclined ($i>30\degr$) galaxy has been multiplied by a factor $f_{\rm
inc}$ as reported in Tab.~\ref{fi}.

\begin{table}
\caption[]{Correction factors, $f_{\rm inc}$, for inclined ($i>30\degr$)
spiral galaxies} \label{fi}
\begin{tabular}{cccc}
\hline
             & \multicolumn{3}{c}{SN type}\\
\cline{2-4}
galaxy type  & ~~~Ia~~~   &   ~~~Ib~~~    & ~~~II~~~ \\
\hline
 S0a-Sb  & 1    &  0.50  &  0.50\\
 Sbc-Sd  & 0.50    &  0.25  &  0.25\\
 Sdm-Sm  & 1    &  0.50  &  0.50\\
\hline
\end{tabular}
\end{table}

It is clear that the correction factors for inclination will significantly
change the final SN rates, expecially in Sbc-Sd galaxies. Therefore, the
uncertainty on the correction factors is a major source of errors, as it will
be discussed in Sect.~\ref{corr:err},

It is worth noting that there is no general agreement on the values of the
inclination factors since they depend on the survey type. For
instance, the visual search of Evans is less affected (vdB\&T). The
present values are similar (although not identical) to those adopted in the
past for the Asiago search (CT88); we verified that the differences is in a
different grouping of galaxy and SN types. Finally, we note that the values of
Tab. \ref{fi} resemble the dependence on galaxy and on SN type adopted in
vdB\&T.

\section{Dependence on the galaxy luminosity}
It was first pointed out by Tammann (\cite{tammann:70}, \cite{tammann:74}) that
the SN rate in Sc galaxies is proportional to the galaxy B luminosity. A
similar
relation has also been proved for S0 (CT88).

To test this effect, we calculated the absolute luminosity $L_{\rm B}$, for
each galaxy of the RC3 sample, from the total $B_{\rm T}$ magnitude corrected
for galactic and internal absorption (adopting $M^{\rm B}_\odot = 5.48$ and $H
= 75\,km\,s^{-1}\,Mpc^{-1}$). The blue luminosity was used as reference because
$B$ magnitudes have been measured for a larger number of galaxies. However, it
must be stressed that the near and the far infrared luminosities are better
indicators of star formation than the blue luminosities (Gallagher et al.,
\cite{gall}) and may be physically more closely related to the rate of SNe with
massive progenitors (II, Ib).

\begin{figure*}
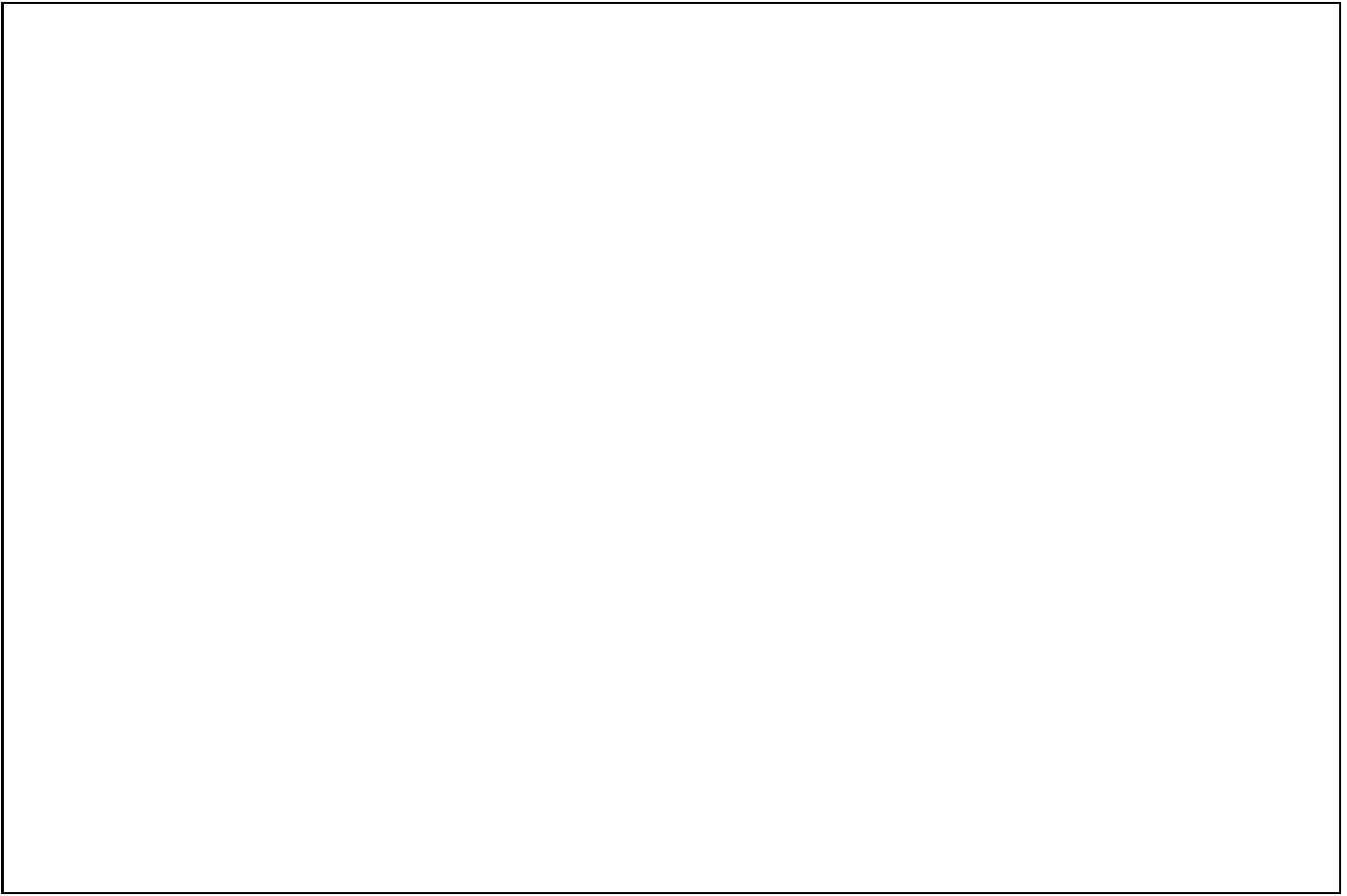

\picplace{12cm}
\caption[]{Dependence of the SN rate on the galaxy B luminosity. In the upper
panels the overall SN rate and the rates of Ia and II, divided in three
luminosity bins are shown. In the lower panels data for different morphological
types of galaxies are plotted. Due to the poor statistics they are split
only into two luminosity bins. In all figures the line passing
through the origin and the barycentre of the points is drawn. In brackets are
the numbers of SNe in each sample.
Indicative error bars are computed assuming Poisson statistics for the SN
events.  }
\label{lum}
\end{figure*}

After correcting for the overexposure and inclination biases, as described in
previous sections, we derived the SN rates for the different SN and galaxy
types
binned according to the galaxy luminosity. The results are summarized in
Fig.~\ref{lum}, where indicative error bars are also reported.

The figure clearly shows that the SN rate is proportional to the B luminosity.
This is true for all SN types (even if we could not test for type Ib due to the
small statistics) and along the Hubble sequence of galaxy morphological types.

These results confirm previous findings and make it possible to compute SN
rates in the so-called {\em SN units}, i.e. $1\,SNu =
1\,SN\,(100\,yr)^{-1}\,(10^{10}\,L_{\rm B\,{\sun}})^{-1}$. We remind
the reader that SN rates expressed in SNu depend on the adopted distance scale,
i.e. in our case they scale as $(H/75)^2$.

\begin{table}
\caption[]{Dependence of SN rate on galaxy catalogue parameters.}\label{tlum}
\begin{tabular}{cccccc}
\hline
          & $<L_{\rm B}^0>$ & \multicolumn{4}{c}{SN rates [SNu]}\\
\cline{3-6}
          & [$10^{10}L_\odot$]& E-S0 & S0a-Sb & Sbc-Sd & All\\
\hline
Tully     &  1.69 &  0.13 & 0.83 & 3.00 & 0.95\\
RSA       &  2.83 &  0.12 & 0.42 & 1.80 & 0.66\\
RC3       &  2.11 &  0.11 & 0.60 & 2.17 & 0.75\\
\hline
\end{tabular}
\end{table}

\section{Dependence on galaxy catalogue parameters}
Using the galaxies in common between three different catalogues we showed in
Sect.8 of Paper I that discrepant morphological classifications induce
significant differences in the SN rates computed  for specific galaxy types. In
particular, the rate of SNe in a galaxy type critically depends on the
morphological classification of the few SN host galaxies, rather than on the
existing systematic differences between the catalogues. Once the frequency is
computed for groups of morphological types or for a catalogue in full, the
differences on the SN rates are washed out.

We now want to test the effects due to the discrepancies in the
magnitudes and absorption corrections reported by different catalogues.
With this aim in mind we calculated the
SN rates in SNu for the galaxies in the Tully (\cite{tully}), RSA
(Sandage \& Tammann \cite{rsa})  and RC3 (instead of RC2 as in Paper I)
catalogues. The
common sample contains 456 galaxies with 39 SNe.
To compute the SN rates we used for each galaxy the distance,
morphological type, magnitude, galactic and internal absorption, taken from
each of the three catalogues. We note that
while some quantities do not change significantly passing from one catalogue to
another, e.g. apparent magnitude and distance, important systematic differences
do occur for the internal and galactic extinctions. The results are reported in
Tab. \ref{tlum}. From the analysis of the common galaxy sample, we reached the
following conclusions:

\begin{enumerate}
\item There is a systematic difference among the corrected luminosity $L_{\rm
B}^0$ reported in the three catalogues, mostly due to the different corrections
adopted for the Galactic and internal absorptions (col.~2 of Tab.~3). In
particular, the average RC3 luminosity is $\sim 1.25$ times that of Tully and
$\sim 0.75$ times that of the RSA. This reflects directly on the SN rates
measured in SNu (cols. 3-6) and must be kept in mind when comparing results of
authors using different source catalogues.

\item Random errors and/or systematic differences in the morphological
types reported by the catalogues cause some variations in the relative
rates of SNe. Obviously, the uncertainty is, more severe when a single
morphological class is considered. For instance, we have verified that the SN
rate in Sc galaxies of Tully is $\sim 4$ times the average value computed for
all morphological types, while the same ratio is only $\sim 2.7$ if data are
taken from RSA. Some effects persists even when galaxies are grouped on a less
detailed basis: e.g. the ratio of SN rates in Sbc-Sd and E-S0 galaxies is
$\sim15$ using the RSA classification and $\sim23$ with the Tully morphological
classification.
\end{enumerate}

\begin{table*}
\caption[]{SN rate per unit blue luminosity in different types of RC3 galaxies
($H=75$ km s$^{-1}$ Mpc$^{-1}$)}
\label{type}
 \begin{flushleft}
\begin{tabular}{crcrrrrrrrr}
\hline
\multicolumn{2}{c}{galaxy} && \multicolumn{3}{c}{SNe}&&\multicolumn{4}{c}{SN
rate [SNu]} \\
\cline{1-2}\cline{4-6}\cline{8-11}
type   & num. && Ia~ & Ib~ & II~ && Ia~   & Ib~  & II~ & All~~~~ \\
\hline
E      &  263&& 3.0&    &     && 0.11 &     &     & 0.11$\pm$0.06\\
S0     &  437&& 4.0&    &     && 0.15 &     &     & 0.15$\pm$0.08\\
S0a,Sa &  274&& 5.2& 0.8& 1.0 && 0.30 & 0.15& 0.19& 0.64$\pm$0.24\\
Sab,Sb &  432&& 5.0& 1.2& 3.8 && 0.12 & 0.12& 0.36& 0.60$\pm$0.19\\
Sbc    &  202&& 2.8& 0.4& 4.8 && 0.22 & 0.10& 1.19& 1.51$\pm$0.53\\
Sc     &  173&& 5.9& 2.7& 6.4 && 0.50 & 0.54& 1.45& 2.49$\pm$0.64\\
Scd,Sd &  351&& 5.3& 0.3& 6.4 && 0.48 & 0.09& 1.87& 2.43$\pm$0.70\\
Sdm-Im &  304&& 1.5& 1.1& 1.4 && 0.20 & 0.30& 0.40& 0.90$\pm$0.45\\
\\
E,S0   &  700&& 7.0&    &     && 0.13 &     &     & 0.13$\pm$0.05\\
S0a-Sb &  706&& 10.2& 2.0& 4.8 && 0.17 & 0.13& 0.30& 0.60$\pm$0.15\\
Sbc-Sd &  726&&14.0& 3.4&17.6 && 0.39 & 0.27& 1.48& 2.14$\pm$0.36\\
\\
All($^*$) & 2461&&34.6& 6.6& 23.8&& 0.22 & 0.11& 0.41& 0.74$\pm$0.09\\
\hline
\end{tabular}
\end{flushleft}

($^*$) including 8 I0 with 1 SNIa and 17 Pec with 1 SNIa.
\end{table*}

\section{SN rates per unit luminosity}\label{snu}
\begin{figure}
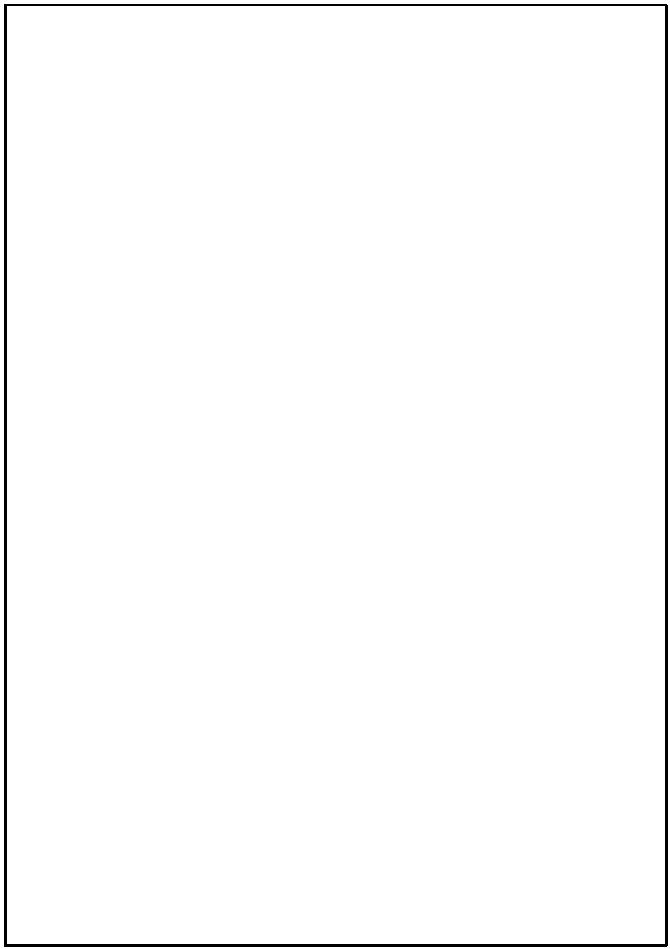

\picplace{12.5cm}
\caption[]{SN rates per unit blue luminosity for galaxies of different
morphological types. The 1-standard deviation Poisson errors are also reported.
Clearly, when a single morphological type is considered, statistical errors
becomes quite large, expecially for SNIb whose discoveries are less frequent.
SN rates in
SNu scale as $(H/75)^2$.} \label{type:fig}
\end{figure}

In Tab.~\ref{type} and Fig.~\ref{type:fig} we report the SN rate per unit blue
luminosity for the different types of galaxies and SNe (based on the RC3
sample). Because of the small numbers, IIP and IIL have been lumped together
and unclassified SNe (cf. Table 2 of Paper I) have been distributed among
different subtypes with the following prescriptions: {\em a)} only Ia are
allowed in E-S0 or I0 galaxies (Barbon et al. \cite{cat}); {\em b)} the average
{\em ratio of discovery} of our searches among the different SN subclasses is
Ia = 4$\times$Ib, IIP=IIL and I=0.6$\times$All. In Tab.~\ref{type}, the
numbers of galaxies are listed in col. 2, the numbers of SNe in cols.~3--5,
while the rates for the different SN types are in cols.~6--8. In the
last column are reported the total SN rates and the errors due only to the
statistic of these events.

{}From Tab.~\ref{type} and Fig.~\ref{type:fig} it is clearly seen that the SN
rate is strongly dependent on Hubble type, with late--spirals (Sbc-Sd) being
about 15 times more prolific than early--type galaxies (E-S0). This is mainly
due to the contribution of type II and Ib SNe. However, it is important to note
that SN~Ia are also $\sim3$  times more frequent in late spirals than they
are in ellipticals.

For Ib/c the statistics are very poor, and, consequently, the uncertainty
in the supernova rates is large. The ratio of the rate of Ib and Ia supernovae
in late spirals is $\nu(Ib)/\nu(Ia)=0.7$. Even allowing for the severe
uncertainties in the adopted absolute magnitude and light curve of Ib
(discussed in Paper I), the above ratio reaches, in the most unfavourable case
(cf. Tab.~\ref{errors}), the value of 2. Our result is in good agreement with
the same conclusion by Evans et al. (\cite{evans:etal}) and vdB\&T, but
significantly different from $\nu(Ib)/\nu(Ia)\sim4$ found by Muller et al.
(\cite{muller:etal}) (cf. Sect. \ref{dis}).

Some concerns may arise because in Table~\ref{type} we redistributed
unclassified SNe among the known subtypes using an empirical rule. To verify
the relative ratio between different subtypes (without considering the absolute
values of the frequencies), we report in Tab.~\ref{late} the relative rates of
occurrence of SNe in late spirals that were computed using only classified SNe.
We confirm that, within the errors, Ia and Ib have similar rates whereas SNII
occur at a significantly higher rate. Moreover, we show that the rates of
IIP and IIL SNe is very similar.

\begin{table}
\caption[]{Relative rates [$\nu ' = \nu(SN~type)/\nu(all)$] of SNe
in late spirals (Sbc-Sd) computed only with classified SNe} \label{late}
\begin{flushleft}
\begin{tabular}{lcccc}
\hline
SNe    & Ia & Ib  & IIP & IIL \\
\hline
num. &8  & 2  & 7  & 6  \\
$\nu'$ & $0.15\pm0.05$ & $0.10\pm0.07$ & $0.39\pm0.15$ & $0.36\pm0.15$ \\
\hline
\end{tabular}
\end{flushleft}
\end{table}

\subsection{The rate of SNe similar to 1987A} \label{87a}
After the discovery of the unusual SN 1987A in the LMC it has been claimed that
this kind of faint SNII  may be the most common SN event; the discovery of this
type of SNe being strongly biased by low intrinsic luminosity (Schmitz \&
Gaskell \cite{sch:gas}). It has also been proposed that some of the faint
(M$<-14$) SNII discovered in the past are similar to SN~1987A.
In particular SN 1973R (in the Sb galaxy NGC3627) and 1982F (in the Sd galaxy
NGC4490) have been mentioned (van den Bergh \& McClure \cite{vdb:mcc:89}).

In analogy to what was done for the other SN types, we computed the integral
control time of 1987A-like SNe for each galaxy type. Because SN~1987A is
not in our SN sample, our derived rate for this SN subtype is
$\nu=0$. If we admit that the two afore mentioned SNe (belonging to our SN
sample) were similar to SN~1987A, we can derive the rates of 1987A-like SNe
$\nu({\rm S0a-Sb})=0.35$ and $\nu({\rm Sbc-Sd})=0.48$ SNu, consistent with the
average value of van den Bergh \& McClure (\cite{vdb:mcc:89}). In this case,
the two SNe have to be removed from the sample of other SNII and the
frequencies of SNII (Plateau +Linear) in spirals become 0.77 SNu, which is
almost twice the estimate for 1987A-like SNe (0.40 SNu).

It must be noted, however, that the light curves of the two SNe were rather
normal plateau--type while the color indices were very red (Patat et al.
\cite{patat}). It is, therefore, likely that they suffered a heavy reddening
and that they were intrinsically different form SN 1987A. Therefore, the
reported faint SNII rate has to be considered an upper limits.

\section{Errors}

\subsection{Influence of the correction factors} \label{corr:err}
We already noted that the correction factors discussed in Sect.\ref{corr:sec}
have a major influence on the final SN rates and that the uncertainties on
their values affect not only the absolute values of the frequencies, but
also the relative ratios among the SN types and among galaxy types. For
instance, it may be argued that the peak of the SN rate in Sbc-Sd
galaxies, $\sim4$ times than in S0a-Sb, is not intrinsic but due to an
overestimate of the adopted correction factor for the late spirals (cf. Sect.
\ref{secinc}).

For the sake of comparison, we show in Table~\ref{corr} the SN rates computed
without correction factors (cols. 2--5)  and with the factors adopted in
Sect.~\ref{corr:sec} (cols. 6--9).
The corrections increase the SN rates by about 40\% for E-S0,
$\sim88$\% for S0a-Sb, $\sim210$\% for Sbc-Sd and the overall SN rates by
$\sim85\%$. We also tested on the RC3 sample the correction factors used in
CT88 and found SN rates $\sim7\%$ larger than those derived with the new
factors.

\begin{table}
\caption[]{Influence of the correction factors on SN rates}\label{corr}
\begin{flushleft}
\begin{tabular}{cccccccccc}
\hline
     &     \multicolumn{9}{c}{SN rate [SNu]}\\
\cline{2-10}
galaxy & \multicolumn{4}{c}{plain} & & \multicolumn{4}{c}{corrected}  \\
\cline{2-5}\cline{7-10}
type & Ia  & Ib & II & all & & Ia  & Ib & II & all \\
\hline
E-S0   & 0.09 & 0.00 & 0.00 & 0.09 && 0.13 & 0.00 & 0.00 & 0.13 \\
S0a-Sb & 0.13 & 0.06 & 0.13 & 0.32 && 0.17 & 0.13 & 0.30 & 0.60 \\
Sbc-Sd & 0.18 & 0.08 & 0.43 & 0.69 && 0.39 & 0.27 & 1.48 & 2.14 \\
\\
All    & 0.14 & 0.06 & 0.20 & 0.40 && 0.22 & 0.11 & 0.41 & 0.74 \\
\hline
\end{tabular}
\end{flushleft}
\end{table}

To estimate the errors induced by these factors, we computed the SN rates
considering a 50\% uncertainty on the number of SNe lost by overexposure and
by inclination. The resulting errors are reported in cols. 11--13 of
Tab.~\ref{errors}, along with the errors from other sources as described in the
next section.

It is clear that this source of error is very important for all galaxies
except E-S0, for which no inclination is involved.

\subsection{Cumulative errors}
Following the above discussion and the tests performed in Paper I, we can now
give reasonable estimates of the total errors on SN rates. In Tab.~\ref{errors}
we list, along with the SN rates (cols. 2--4), the contributions of the
different sources of error: in cols. 5--7 those derived assuming  Poisson
statistics for the SN events, in cols. 8--10 those related to the uncertainties
in the input parameters (cf. Tab.~4 of Paper I), in cols. 11-13 those due to
the uncertainties on the correction factors described in the previous section.
Finally in cols. 14--16 are reported the cumulative errors obtained adding,
in quadrature, the various terms. While not formally correct, given the
different meaning of the individual items, we think that this numbers is
a reasonable estimates of the total uncertainties of SN rates.

Of course, when considering a particular SN type in a given type of galaxy, the
determinations of the SN rates are based on small numbers of SNe and,
therefore, the errors associated with the statistics of the events dominate.
However, if we bin the galaxies in wider morphological classes, as in
Tab.~\ref{errors}, then the contribution of the statistics to the total error
is reduced. The uncertainties in input parameters, discussed in Paper I, give
in all cases a small contributions. Instead, for spirals the uncertainty due to
the correction factors, in particular the contribution of the  inclination
along the line of sight, is dominant. Ellipticals do not have such term,
so that statistical errors prevail.
Hence, the actual total errors of the SN rates are significant larger than
those due to the statistics alone, which dominate only when the SN sample
is small.

\begin{table*}
\caption[]{Different contributions to the errors of the SN rates [SNu]}
\label{errors}
\begin{flushleft}
\begin{tabular}{crrr|rrrrrrrrrrrrrrrr}
\hline
galaxy & \multicolumn{3}{c|}{SN rate}
&&\multicolumn{3}{c}{SN statistics} &&
\multicolumn{3}{c}{input parameters} &&\multicolumn{3}{c}{correction factors}
&& \multicolumn{3}{c}{total errors}\\
\cline{2-4}\cline{6-8}\cline{10-12}\cline{14-16}\cline{18-20}
type   &  Ia & Ib   & II  && Ia   & Ib  & II  && Ia  & Ib  & II  &&
Ia  & Ib & II && Ia   & Ib   & II \\
\hline
 &&&&  &      &     &     &      &     &     &     &    &    &     &    &    &
     &     &     \\
E-S0   & 0.13 &     &     && 0.05 &     &     &&0.02&    &    && 0.02&    &
&
&0.06 &      &     \\
S0a-Sb & 0.17 & 0.13& 0.30&&0.05 & 0.09& 0.14&& 0.02&0.03&0.05&& 0.05&
0.05&0.12&& 0.07 & 0.11 & 0.19\\
Sbc-Sd & 0.39 & 0.27& 1.48&& 0.10 & 0.15& 0.35&& 0.05&0.05&0.26&& 0.15&
0.09&0.48&& 0.19 & 0.18 & 0.65\\
&&&&  &      &     &     &      &     &     &     &    &    &     &    &    &
     &     &     \\
\hline
\end{tabular}
\end{flushleft}
\end{table*}

\section{Discussion}\label{dis}
In a recent paper vdB\&T carefully reviewed the determinations of the SN rates
published by Tammann (\cite{tammann:82}), Evans et al (\cite{evans:etal}) and
CT88. After applying normalization factors to each set of statistics, they
derived their best estimates of the SN rates (cf. their Tab. 8).

A direct comparison between the result of that paper and our new estimates is
not straightforward, because they refer to $L_B$ luminosities {\em not
corrected for (large) internal absorption or for inclination}, whereas we used
corrected $L_{\rm B}^0$ luminosities. A test on our sample of late spirals
shows that the SN rate expressed in SNu decreases by a factor 2
if one uses $L_{\rm
B}^0$ instead of the uncorrected luminosities. Moreover, because the
luminosities of E-S0 change less (being not affected by internal absorption)
this also alters the relative SN rates between early and late Hubble types.
Keeping this in mind, we compare our revised SN rates with those of vdB\&T,
scaling the results to our adopted value of the Hubble constant.

First of all, our rate of SN~Ia in E-S0 galaxies, $\nu_{Ia}=0.13\pm0.06$ SNu,
is a factor 4 smaller than that in vdB\&T. Moreover, our $\nu_{Ia}$ in E-S0 is
1/3 of that in late spirals, while is $\nu_{Ia}({\rm E-S0}) \sim 2\times
\nu_{Ia}({\rm Sbc-Sd})$ in vdB\&T, who assumed the rate of SNIa constant from
early to late spirals. Such large discrepancies may be partially due to the
different sources of the galaxy parameters, in particular, the luminosity and
the internal absorption, and to the poor statistics in this type of galaxies.

Alternatively, the adoption of different corrections for selection
effects can play a role.
In principle, we might have overestimated the selection effects in spirals and
simultaneously neglected possible selection effects in E-S0
galaxies. For instance, the {\em shading} of SNe in the nuclear region of
galaxies might be more severe in E than in Sc galaxies. However, if the rates
of SN~Ia of vdB\&T are correct then our searches must have missed almost 75\%
of the SNe in E galaxies, which is hard to believe.
However, our total rates of SNe in late spirals are in fair agreement with the
corresponding value of vdB\&T, after accounting for the difference in
luminosity
scale.

The relative frequency of type II to Ia supernovae in late spirals is
$\sim4$ according to our estimates, and $\sim8$ according to vdB\&T. Again,
this is probably related to their {\em a priori} assumption of a uniform
$\nu_{Ia}$ in all spirals.

We already mentioned in Sect~\ref{snu} that our results contradict the finding
by Muller et al. (\cite{muller:etal}) of a very high rate of SN~Ib/c. As a
possible explanation of this disagreement, one may recall that SN~Ib/c have
been clearly separated from SNIa only after 1985. Since our searches have
produced most of the discoveries earlier,  one might think that there is a bias
in the subclassification within our SNI sample. Whereas this cannot be
completely ruled out, we remind the reader that published spectra have been
examined by Branch (\cite{branch}) to assign, when possible, each SNI to the
appropriate subtype.
In fact, in our SN sample only $\sim30\%$ of SNI could not be better
classified, excluding SNe in E-S0 galaxies that, safely,  can be assumed
to be of type Ia. Moreover, from an updated version of the Asiago SN catalogue
(Barbon et al. \cite{cat}), we found that the percentage of SNIb among the
classified SNI is $\sim20\%$. This is so if we consider the SNe discovered in
the most active period of our searches (1960-1988), but also during
the following 4 years.

On the other side, whereas the Berkeley CCD search independently discovered
$\sim15\%$ of all SNe announced after 1986 (Muller et al. \cite{muller:etal}),
they found $\sim45\%$ of all SNe classified Ib/c in that period. This can
probably be understood considering that CCDs have a  very high red sensitivity,
compared with normal photographic plates, and that at maximum SN~Ib are
significantly redder ($(B-V)_0\sim 1$ and $(V-R)_0\sim 0.5$) than SN~Ia and II
($(B-V)_0\sim 0$ and $(V-R)_0 \sim 0$). Instead, the SN maximum
luminosities adopted by Muller et al. (\cite{muller:etal}) for their frequency
computations are more suited for  blue sensitive detectors. This is not a
serious problem for Ia and II, given their average colors, but make a big
difference for Ib's that can be as luminous as SNeIa in the yellow--red
spectral region. If this is true, the SNIb rate of Muller et al
(\cite{muller:etal}) should be reduced by a factor $\sim 2$ (cf. Paper I).
This also reduces the disagreement with our Ib rate.

\subsection{The local SN rate}\label{local}
{}From the values reported in Tab.~\ref{type}, we can determine the expected SN
rates in our Galaxy and compare them with values obtained from other sources.
We
assume the Galaxy to be of type Sb$\pm0.5$ and to have a total luminosity
$L_{\rm B} = 2.0\pm0.6 \times 10^{10}\,L_\odot$ (van der Kruit \cite{kruit}).
In
order to reduce statistical uncertainties, we use the average SN rates for the
galaxy types Sab-Sbc. With these assumptions, the Galaxy
is expected to produce $3\pm2$ Ia, $2\pm2$ Ib, and $12\pm8$ II, i.e.
$17\pm9$ SNe {\em per millennium} (considering only the uncertainties due to SN
statistics and to the luminosity of the Galaxy).

Six historical supernovae are known to have exploded in the Milky Way in the
last millennium, but many more have been lost due to the heavy absorption in
the galactic disk. The evaluation of the percentage of obscured SNe is not
straightforward, however different authors (cf. vdB\&T) agree in giving
galactic SN rates ($>50$ SNe per millennium) larger than the results discussed
above. The disagreement may be reduced if the Galaxy has in fact a higher
luminosity or if one adopts a later Galactic type.

The SN rate in the Galaxy can also be derived  from the observed distribution
of radio SN remnants. The result is $3.4\pm2.0$ SNe per century (vdB\&T).
Within the errors this is consistent with our estimate ($1.7\pm0.9$). Also note
that our value is in very good agreement with the {\em best estimate} of the
galactic rate of $\sim2$ SNe per century  derived by van den Bergh
(\cite{vdb:90}), analysing galactic SNR's, historical SNe, and novae in M31 and
M33.

Finally, it is interesting to compute the total expected number of SNe per
century in the galaxies of the Local Group, where it is reasonable to assume
that no SN has been missed during the last century. Using the SN rates (in SNu)
of Table~\ref{type} and using for each local galaxy the morphological type and
luminosity reported in the RC3, we expect 2.9 SNe per century (excluding our
own Galaxy). This number is in good agreement with the observed number of 2 SN
in the last century (1885A in M31 and 1987A in LMC).

\subsection{Comparison with theoretical predictions}
Theoretical estimates of the SN rates have been derived from models for
stellar and galaxy evolution. There are, in principle, many different
explosion mechanisms that can led to a SN event, either from massive single
star progenitors or from binaries. Tutukov et al. (\cite{tutu:etal}) proposed,
depending on different properties of binary systems,  two dozen possible kinds
of SNe, many of which still remain without an observed counterpart. Using a
numerical program they calculated the expected numbers per century of different
SN types in the Galaxy. The results can be lumped in the two major types as
follows: 1.0 to 1.9 SNe per century for type I and 1.96 to 3.35 for type II.
These numbers are consistent with our {\em observed} estimates for the Galaxy,
the lower values being favoured.

SN explosions release large amounts of energy and of heavy elements,
synthesized during the evolution of the progenitors and in the explosions,  in
the interstellar medium. Therefore, they play
a major role in the chemical and dynamical evolution of galaxies.
Tornanb\'e \& Matteucci (\cite{torn}) derived the present rate of SNIa in
ellipticals from an evolutionary model. With the present work, the discrepancy
between their predicted value, 0.04 SNu, and the observations is reduced.
Moreover, they found that the expected rates of SNIa and SNIb in the Galaxy are
almost identical (assuming that SNIb progenitors are in binary systems
consisting of a degenerate dwarf and a non-degenerate He-star).

More recently, Ferrini and Poggianti (\cite{ferrini}) presented a multiphase
model of the evolution of elliptical galaxies that includes the chemical
galactic enrichment and gas removal due to SN explosions. Their rate
of SN~Ia in E is in the range 0.01 to 0.10 SNu, consistent with the observed
rate of Tab.~\ref{type}.

Models of the chemical evolution for the solar neighbourhood and the whole
disk, including detailed nucleosynthesis from SNe, have been computed by
Matteucci \& Fran\c{c}ois (\cite{matt:fra}). They  predict for our Galaxy 0.4
SNIa per century, 0.4 SNIb and 1.1 SNII, in remarkable agreement
with our estimates.

\section{Conclusions}
A significant enlargement of the galaxy and of the SN samples used for the
determination of the frequency of SNe with the so-called control time method,
was obtained by including in a common database the observations of two long
term
SN search programmes. This allowed us to estimate the rates of SNe of various
subtypes in different morphological classes of the parent galaxies. In
particular, SNe ia and Ib were considered separately as were SNe IIP and
IIL in late spirals. Also an upper limit to the frequency of SNII similar to
SN 1987A was computed.

We discussed the importance in our combined searches of two major
selection effects, which bias our samples,
against the discovery of SNe due to plate overexposure and
galaxy inclination. The overexposure effect increases with the recession
velocity of the parent galaxies and is stronger in our photographic searches
than in visual and CCD surveys. Nevertheless, contrary to previous claims, even
the latter surveys are affected when more distant galaxies are considered. The
effect of galaxy inclination is large and depends on the galaxy morphology and
SN type (cf. Tab. \ref{fi}).

The rates of SNe in different galaxy types (cf. Tab.~\ref{type}) have been
computed in $SNu = 1\,SN\,(100yr)^{-1}\,10^{10} L_{\rm B})^{-1}$, after proving
that the rates of occurrence of all SN types is proportional to the luminosity
of the host galaxies. The derived rate of SNe (Ia) in E galaxies
is smaller than in other
galaxy types and is considerably reduced with respect to previous
determinations (e.g. vdB\&T). The rate of SNeIb in spirals is found to be
similar to that of SNe Ia while II are the most frequent SN type.

A particular effort has been devoted to a reliable determination of the errors
of the SN rates originating from the input assumptions on the SN
parameters (Paper I) and from the correction factors discussed in Sect.
\ref{corr:sec}. While the errors of the SN rate in ellipticals is still
dominated by the SN statistics, in spirals the uncertainties in the correction
factors is dominant. In general, we show that the total errors of the SN
rates are larger than those usually reported in the literature. We also tested
the effect on the final SN rates of the parameters of the sample galaxies
reported in three different catalogues and found that the adoption of a
different correction for galactic and internal absorption will modify the SN
rates (in SNu) and makes it hard to compare supernova rate determinations based
on different source catalogues.

Accounting for all the sources of uncertainties our best estimates of the {\em
absolute} SN rates are in the following ranges: 0.05 to 0.2 in E-S0,
0.2 to 1 in S0a-Sb and 1 to 3.5 SNu in Sbc-Sd.


\begin{thebibliography}{cosaservira}     % (do not forget {})
\bibitem[1989]{cat}
   Barbon, R., Cappellaro, E., Turatto, M., 1989, A\&AS 81, 421
\bibitem[1991]{bart:etal}
   Bartunov, O. S., Makarova, I. N., Tsvetkov, D. Yu., 1991,
   Astron. Zh. Pis'ma, 17, 164.
\bibitem[1990]{branch}
   Branch, D., 1990, in {\em Supernovae} ed. Petschek, A.G., Springer-Verlag,
	New York, p.30
\bibitem[1988]{capp:tur}
   Cappellaro, E., Turatto, M., 1988, A\&A 190, 10 ({\bf CT88})
\bibitem[1992]{paper1}
    Cappellaro, E., Turatto, M., Benetti, S., Tsvetkov, D.Yu., Bartunov, O.S.,
    Makarova, I.N. 1992 A\&A in press
\bibitem[1991]{rc3}
   de Vaucouleurs, G., de Vaucouleurs, A., Corwin, H. G., Buta, R.J.,
   Paturel, G., Foque, P., 1991,
   Third Reference Catalogue of Bright Galaxies, Springer-Verlag (New York)
    ({\bf RC3})
\bibitem[1989]{evans:etal}
   Evans, R., van den Bergh, S., McClure, R. D., 1989, ApJ 345, 752
\bibitem[1992]{ferrini}
   Ferrini, F., Poggianti, B.M., 1992, ApJ in press
\bibitem[1984]{gall}
    Gallagher III, J.S., Hunter, D.A., Tutukov, A.V., 1984, ApJ 345, 752
\bibitem[1990]{gut}
    Guthrie, B.N.G., 1990, A\&A 234, 84
\bibitem[1989]{matt:fra}
    Matteucci, F., Fran\c{c}ois, P., 1989, MNRAS 239, 885
\bibitem[1992]{muller:etal}
    Muller, R.A., Marvin, H.J., Pennypacker, C.R., Perlmutter, S.,
    Sasseen, T.P., Smith, C.K. 1992 ApJ 384, L9
\bibitem[1992]{patat}
	Patat, F., Barbon, R., Cappellaro, E., Turatto, M., 1992, A\&AS
	in press
\bibitem[1981]{rsa}
   Sandage, A., Tammann, G. A., 1981, A Revised Shapley-Ames Catalog of
   Bright Galaxies, Carnegie Institution, Washington ({\bf RSA})
\bibitem[1988]{sch:gas}
     Schmitz,M.,F. Gaskell, C.M., 1988, in {\em Supernova 1987A in the Large
     Magellanic Cloud} eds. Kafatos, M. and Michalitsianos, A.G. Cambridge
     University Press  (Cambridge)
\bibitem[1979]{shaw}
    Shaw, R.L., 1979, A\&A 76, 188
\bibitem[1970]{tammann:70}
   Tammann, G. A., 1970, A\&A 8, 458
\bibitem[1974]{tammann:74}
   Tammann, G. A., 1974, in {\em Supernovae and Supernova Remnants}, ed. C.B.
   Cosmovici, Reidel, Dordrecht, p.\ 155
\bibitem[1977]{tammann:77}
   Tammann, G. A., 1977, in {\em Supernovae}, ed. D.N Schramm, Reidel,
   Dordrecht, p.\ 95
\bibitem[1982]{tammann:82}
   Tammann, G. A., 1982, in {\em Supernovae: A Survey of Current Research},
eds.
   M. J. Rees, R. J. Stonehouse, Reidel, Dordrecht, p.\ 371
\bibitem[1987]{torn}
	Tornanb\'e, A., Matteucci, F., 1987, ApJ 318, L25
\bibitem[1988]{tully}
   Tully, R. B., 1988, {\em Nearby Galaxies Catalog}, Cambridge University
   Press, Cambridge
\bibitem[1992]{tutu:etal}
    Tutukov, A. V., Yungelson,L. R., Iben, I. Jr., 1992, ApJ 386, 197
\bibitem[1990]{vdb:90}
    van den Bergh, S., 1990, in {\em Supernovae} ed. S.E. Woosley,
    Springer-Verlag,New York
\bibitem[1989]{vdb:mcc:89}
    van den Bergh, S., McClure, R.D. 1989, ApJ 347, L29
\bibitem[1990]{vdb:mcc}
    van den Bergh, S., McClure, R.D. 1990, ApJ 359, 277
\bibitem[1991]{vdb:tamm}
    van den Bergh, S., Tammann, G. A., 1991, ARA\&A 29, 363 ({\bf vdB\&T})
\bibitem[1987]{kruit}
     van den Kruit, P.C., 1987, in {\em The Galaxy} eds. G. Gilmore, B.
Carswell
     D. Reidel Publishing Company (Dordrecht)
\end{thebibliography}
\end{document}